\documentclass[12pt,preprint]{aastex}

\newcommand{\be}{\begin{equation}}
\newcommand{\ee}{\end{equation}}

\begin{document}

\title{Non-Gaussianity of the Cosmic Baryon Fluid: Log-Poisson
   Hierarchy Model}

\author{Ji-Ren Liu and Li-Zhi Fang}

\affil{Department of Physics, University of Arizona, Tucson, AZ
85721}

\begin{abstract}

In the nonlinear regime of cosmic clustering, the mass density field
of the cosmic baryon fluid is highly non-Gaussian. It shows
different dynamical behavior from collisionless dark matter.
Nevertheless, the evolved field of baryon fluid is scale-covariant
in the range from the Jeans length to a few ten h$^{-1}$ Mpc, in
which the dynamical equations and initial perturbations are scale
free. We show that in the scale-free range, the non-Gaussian
features of the cosmic baryon fluid, governed by the Navier-Stokes
equation in an expanding universe, can be well described by a
log-Poisson hierarchical cascade. The log-Poisson scheme is a random
multiplicative process (RMP), which causes non-Gaussianity and
intermittency even when the original field is Gaussian. The
log-Poisson RMP contains two dimensionless parameters: $\beta$ for
the intermittency and $\gamma$ for the most singular structure. All
the predictions given by the log-Poisson RMP model, including the
hierarchical relation, the order dependence of the intermittent
exponent, the moments, and the scale-scale correlation, are in good
agreement with the results given by hydrodynamic simulations of the
standard cold dark matter model. The intermittent parameter $\beta$
decreases slightly at low redshift and indicates that the density
field of baryon fluid contains more singular structures at lower
redshifts. The applicability of the model is addressed.

\end{abstract}


\section{Introduction.}

In the universe, about 72\% of the energy density is in the form of
dark energy, 24\% cold dark matter, and a small fraction, 4\% baryon
matter. Since the dark energy is assumed to be spatially uniform,
the dynamics of the clustering of cosmic baryon fluid should be
dominated by the underlying gravitational potential of dark matter.
However, it has already been recognized in the early study of cosmic
structure formation that in the nonlinear regime the dynamical
behavior of the cosmic baryon fluid, or of the intergalactic medium
(IGM), doesn't always follow the collisionless dark matter. Although
the cosmic baryon fluid is passive substance in comparing with dark
matter, it statistically decouples from the underlying dark matter
field in the non-linear evolutionary stage. In the scale free range,
the cosmic baryon fluid, as a Navier-Stokes fluid in the expanding
universe, is similar to the fluid being moved by inertia, and should
show some features as the turbulence in inertial range (Shandarin
and Zeldovich 1989).

Later, it was found that the dynamical equations of the velocity
fields of cosmic matter essentially are a variant of the
random-force-driven Burgers' equation (Gurbatov et al. 1989; Berera
\& Fang 1994). For baryon fluid, it is a Burgers' equation driven by
the random force of the gravity of dark matter (Jones 1999;
Matarrese \& Mohayaee 2002). Burgers' fluid will show highly
non-Gaussian features due to the development of Bergers' turbulence
when the Reynolds number is large enough (Polyakov 1995; L\"assig
2000; Bec \& Frisch 2000; Davoudi et al. 2001). In this state, the
Burgers' fluid consists of shock waves in low as well as in high
density regions, and therefore, non-Gaussianity can be seen in low
as well as in high density regions.

This property has received supports from the absorption spectra of
QSOs, which is caused by the IGM with moderate mass density. For
example, the Ly$\alpha$ transmitted flux in the absorption spectra
of QSOs is found to be significantly intermittent and its
probability distribution functions (PDF) are remarkably long tailed
(Jamkhedkar et al. 2000; Pando et al. 2002; Feng et al. 2003). The
\ion{H}{1} and \ion{He}{2} Ly$\alpha$ absorption lines of QSO HE2347
can not be explained by thermal broadening, but consistent with
turbulence broadening (Zheng et al. 2004; Liu et al. 2006).
Moreover, samples produced by cosmological hydrodynamic simulations
also reveal the statistical decoupling of baryon matter from dark
matter and the non-Gaussian features of Bergers' turbulence (He et
al. 2004, 2005; Kim et al. 2005).

One of the latest developments in this direction is that the random
velocity fields of the cosmic baryon fluid are found to be extremely
well described by She-Leveque's (SL) scaling formula (She \& Leveque
1994) in the scale range from the Jeans length to larger than 10
h$^{-1}$ Mpc (He et al. 2006). The SL scaling formula is believed to
characterize the scaling hierarchy of the non-linear evolution of
Navior-Stokes fluid, like fully developed turbulence (e.g., Frish
1995). Therefore, the dynamical features of the cosmic baryon fluid
in the nonlinear regime are similar to the fully developed
turbulence: it is of scaling hierarchy. It is interesting to note
that the SL formula has also been successfully applied to describe
the mass fields of gas on interstellar scales (Boldyrev et al. 2002;
Padoan et al. 2003).

The SL formula actually is originated from a cascade of log-Poisson
random multiplicative process (RMP), which is related to the hidden
symmetry of the Navier-Stokes equations (Dubrulle 1994; She \&
Waymire 1995; Benzi et al. 1996). This motivate us to investigate
whether the clustering behavior of the mass density field of the
cosmic baryon fluid can be described by the log-Poisson RMP.
Theoretically, this is not trivial, because the cosmic baryon fluid
is compressible and dominated by the gravitational field of dark
matter, while the SL scaling formula originally was proposed to
describe the velocity field of incompressible fluid.

The paper is organized as follows. \S 2 describes the model of the
log-Poisson RMP for the nonlinear evolution of the cosmic baryon fluid.
The predictions and its tests of the log-Poisson RMP model are presented
in \S 3. \S 4 presents briefly the redshift-evolution of the coefficients
of the log-Poisson model. Discussion and conclusion are given in \S 5.

\section{Log-Poisson RMP model}

\subsection{The log-Poisson hierarchy}

The clustering and non-Gaussianity of the cosmic mass density and
velocity fields are usually measured by two and multiple point
correlation functions. To reveal the features of the scaling hierarchy of
the mass density field, however, it is more effective to use
the structure function defined by
%eq1
\begin{equation}
S_p(r)\equiv \langle |\delta\rho_r|^p\rangle,
\end{equation}
where $\delta\rho_{r}= \rho({\bf x+r})-\rho({\bf x})$, $r=|\bf r|$,
$p$ is the order of statistics, and the average $\langle ...\rangle$
is taken over the ensemble of density field. For statistically
isotropic and homogenous random field $\rho({\bf x})$, $S_p(r)$
depends only on $r$.

The difference between the correlation function and structure
function has been analyzed in detail by Monin \& Yaglom (1975). The
variable $\delta\rho_{r}= \rho({\bf x+r})-\rho({\bf x})$ is not
$\delta \rho({\bf x})=\rho({\bf x})-\bar{\rho}$, $\bar{\rho}$ being
the mean of density; the variable $\delta \rho({\bf x})$ can be
larger than $\bar{\rho}$, but cannot be less than $-\bar{\rho}$, and
therefore, for a nonlinear field, the distribution of $\delta
\rho({\bf x})$ generally is skew; however, the distribution of
$\delta\rho_{r}$ is symmetric with respect to positive and negative
$\delta\rho_{r}$ if the field is statistically uniform.

In the scale-free range of the fluid, the structure function as a
function of $r$ can be expressed as a power law
%eq5
\begin{equation}
S_p(r)\propto r^{\xi(p)}.
\end{equation}
For fully developed turbulence of Navier-Stokes fluid, $\xi(p)$ is a
nonlinear function of $p$, i.e. the mass field is intermittent, and
$\xi(p)$ is called intermittent exponent (Frisch 1995). Since the
pioneer work of Kolmogorov (1941), it is believed that the relation of
$\xi(p)$ vs. $p$ is related to the scale-covariance of the dynamical
equations and initial conditions. Since then many hierarchy models
for interpreting $\xi(p)$ have been proposed. Finally the best model
is given by the SL scaling formula (She \& Leveque 1994). It has
been shown that the SL formula is yielded from the Log-Poisson
hierarchy process, which is related to the so-called generalized
scale covariance of the Navier-Stokes equations (Dubrulle 1994).
Therefore, one may expect that the statistical behavior of the mass
field of cosmic baryon matter would also be interpreted by the
log-Poisson random multiplicative processes (RMP).

The log-Poisson RMP assumes that, in the scale-free range, the
variables $|\delta\rho_{r}|$ on different scales $r$ are related
from each other by a statistically hierarchy relation given by
%eq3
\begin{equation}
|\delta\rho_{r_2}| = W_{r_1r_2}|\delta\rho_{r_1}|,
\end{equation}
where
%eq4
\begin{equation}
W_{r_1r_2}=\beta^m (r_1/r_2)^{\gamma},
\end{equation}
which describes how the fluctuation $|\delta\rho_{r_1}|$ on the
larger scale $r_1$ related to fluctuations $|\delta\rho_{r_2}|$ on
the smaller scale $r_2$. In eq.(4), $m$ is a Poisson random variable
with the PDF
%eq5
\begin{equation}
P(m)=\exp(-\lambda_{r_1r_2})\lambda_{r_1r_2}^m/m!.
\end{equation}
To insure the normalization
$\langle W_{r_1r_2} \rangle=1$, where $\langle...\rangle$ is over $m$,
the mean $\lambda_{r_1r_2}$ of the Poisson distribution should be
%eq6
\begin{equation}
\lambda_{r_1r_2}= \gamma[\ln(r_1/r_2)]/(1-\beta).
\end{equation}
It is enough to consider only $|\delta\rho_{r}|$, as the
distribution of positive and negative $\delta\rho_{r}$ is symmetric.

The log-Poisson model of equation (3) depends only on the ratio
$r_1/r_2$, thus, it is scale invariant. The model is
determined by two dimensionless positive parameters: $\beta$ and
$\gamma$, of which the physical meaning will be given below.
Equation (3) relates $\delta\rho_{r}$ on different scales by
multiplying a random factor $W$, and therefore, it is a random
multiplicative process (RMP), which generally yields a non-Gaussian
field even when the field originally to be Gaussian (Pando et al.
1998). For a Gaussian field, variables $\delta\rho_{r_1}$ and
$\delta\rho_{r_2}$ are statistically independent and it requires
$\beta \rightarrow 1$ in equation (4).

\subsection{Parameter $\gamma$ and singular structures}

With the log-Poisson model, the intermittent exponent $\xi(p)$ is
given by (see Appendix)
%eq7
\begin{equation}
\xi(p)=-\gamma[p-(1-\beta^{p})/(1-\beta)].
\end{equation}
For a Gaussian field with scale-free power spectrum, we have
$\xi(p)\propto p$. This indicates again that a Gaussian field requires
$\beta \rightarrow 1$.

Considering a density field containing singular structures at
positions ${\bf x_i}$, we have $\rho({\bf x})\propto |{\bf
x-x_i}|^{-\alpha}$ near ${\bf x_i}$ with $\alpha >0$. The variables
$|\delta \rho_r|^p$ near ${\bf x_i}$ should be $\simeq |r|^{-\alpha
p}$, and therefore $|\delta \rho_r|^{p+1}/|\delta \rho_r|^p \simeq
|r|^{-\alpha}$. Thus, to pick up the singular structures, we define
a statistical tool as
%eq8
\begin{equation}
F_p(r) \equiv S_{p+1}(r)/S_p(r).
\end{equation}
For higher $p$, singular structures have larger contributions to
$S_p(r)$, while for lower $p$, weak-clustering structures have
larger contributions to $S_p(r)$. Therefore $F_p(r)$ measures
clustering structures, which are dominant for the $p$-order
statistics. Obviously, when $p\rightarrow \infty$, $F_p(r)$ should
be dominated by the singular structures, i.e., $F_{\infty} \propto
r^{-\alpha}$.

On the other hand, from equations (2), (7), and (8), one finds
%eq9
\begin{equation}
F_p(r) \propto r^{-\gamma(1-\beta^p)}.
\end{equation}
Since $\beta<1$, we have
%eq10
\begin{equation}
F_{\infty}=\lim_{p\rightarrow \infty} \frac{\langle |\delta
\rho_r|^{p+1}\rangle}{\langle |\delta \rho_r|^p \rangle}\propto
r^{-\gamma}.
\end{equation}
Therefore, the parameter $\gamma$ of the log-Poisson RMP is actually
the power-law index of the mass profile of the most singular
structures. It should be pointed out that the word ``singular'' is
applicable only asymptotically, because we cannot let $r\rightarrow
0$ to pick up the singular structure, as $r$ should be in the
scale-free range.

\subsection{Parameter $\beta$ and intermittency}

As mentioned in last subsection, when $\beta \rightarrow 1$, the
field would be Gaussian, and therefore, equation (9) implies that
for a Gaussian field, $F_p(r)$ would be $r$-independent. On the
other hand, for a utmost intermittent field, which consists only the
most singular structures, and the fluctuations between the singular
structures are zero, $F_p(r)$ should be equal to $r^{-\gamma}$
regardless of $p$. From equation (9), the utmost intermittent field
should have $\beta=0$. Thus, the parameter $\beta$ is to measure the
level of intermittency: non-intermittency corresponds to $\beta=1$,
and the strongest intermittency corresponds to $\beta=0$.

The meaning of $\beta$ can also be seen with the following
hierarchical relation of $F_{p}(r)$
%eq11
\begin{equation}
\frac{F_{p}(r)}{F_{\infty}(r)}= \left
[\frac{F_{p+1}(r)}{F_{\infty}(r)}\right]^{1/\beta},
\end{equation}
which can be derived from equations (9) and (10). Equation (11) is
invariant with respect to a translation in $p$. In deriving eq.(11),
we assume that the proportional coefficient of equation (9) is
$p$-independent. We will show that this assumption is correct.

As mentioned in \S 2.2, the $F_p(r)$ measures the clustered
structures dominating the $p$ order statistics. The smaller the $p$,
the larger the contribution of weak-clustering structures to
$F_p(r)$. Therefore, equation (10) describes the hierarchical
relation between the stronger (or high $p$) and weaker (or low $p$)
clustering. In the scale-free range where
$F_{p+1}(r)/F_{\infty}(r)<1$, we have $F_{p}(r)/F_{\infty}(r)<
F_{p+1}(r)/F_{\infty}(r)$ if $\beta<1$. That is, for an intermittent
field, weak clustering structures are strongly suppressed with
respect to the most singular structures; the smaller the $\beta$,
the stronger the suppression of weak clustering structures.

\section{Non-Gaussianity of the cosmic baryon fluid}

The samples for testing the model of \S 2 are similar to that used
in He et al. (2006), which are given by a hybrid
hydrodynamic/$N$-body simulation, consisting of the WENO algorithm
for baryon fluid and $N$-body simulation for particles of dark
matter (Feng et al. 2004). We now produce samples of 50 $h^{-1}$ Mpc
box, $768^3$ grid, and the cosmological parameters are taken from
the results of WMAP (Spergel et al. 2006). The samples output at
redshifts 0, 1, 2, 3, and 4. We randomly sample 10,000
one-dimensional sub-samples at each redshift.

In order to have a complete description of the density field and
avoid false statistical correlation, the variable
$\delta\rho_{r}$ should be given by a proper decomposition of the
field $\rho_{{\bf x}}$. We will use the decomposition of discrete
wavelet transform (DWT), which is found to be effective to describe
turbulence (e.g., Farge 1992). With the DWT, the variables of mass
density field is given by
%eq12
\begin{equation}
\delta\rho_{r}=\int \rho({\bf x})\psi_{\bf j,l}({\bf x})d{\bf x},
\end{equation}
where $\psi_{\bf j,l}({\bf x})$ is the base of discrete wavelet
transform (e.g., Fang \& Thews 1998). For a one-dimensional sample
of length $L$, the scale index $ j$ is related to the scale $r$ by
$r=L/2^{j}$ and the position index $l$ is for the cell at ${\bf
x}=lL/2^{j}$ to $(l+1)L/2^{j}$. We will use the Harr wavelet to do
the calculation below. We also repeat the calculations with wavelet
Daubechies 4. The non-Gaussian statistical features given by
Daubechies 4 are the same as that of Haar wavelet.

\subsection{PDF and structure function}

%fig1
\begin{figure}[htb]
\center
\includegraphics[scale=1,height=3in]{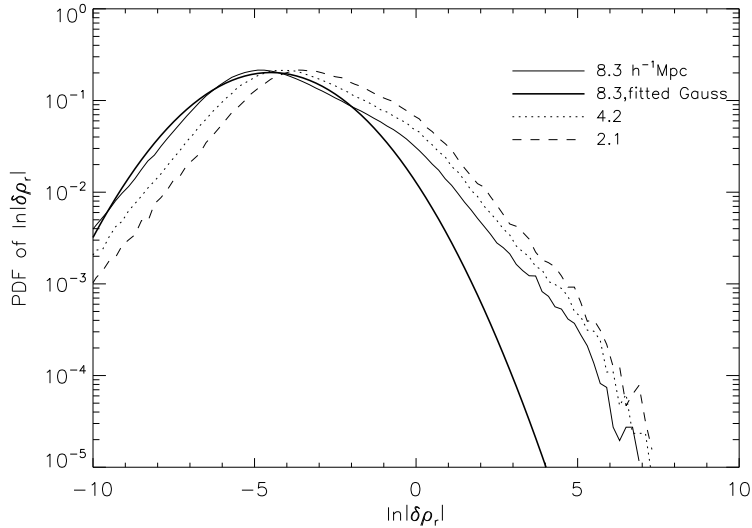}
\caption{PDFs of the density difference $\ln|\delta\rho_{r}|$ with
$r=2.1$, 4.2 and 8.3 h$^{-1}$ Mpc for samples at $z=0$. A fitted
Gaussian PDF of $\ln|\delta\rho_{r}|$ is also shown.}
\end{figure}

We first show the basic statistical deviation of the baryon mass
density field from a Gaussian field. Figure 1 gives the PDF of the
density difference variables,
$p(\ln|\delta\rho_r|)d\ln|\delta\rho_r|$, for the cosmic baryon
fluid sample at $z=0$ on scales $r=2.1$, 4.3, 8.3 $h^{-1}$ Mpc. A
fitted Gaussian to the PDF on scale 8.3 $h^{-1}$Mpc is also shown as
the thick solid line in Figure 1. Since the Gaussian fitting is for
$\ln|\delta\rho_r|$, the fitted curve actually is a lognormal
distribution for $|\delta\rho_r|$. It shows clearly that on all the
scales, the PDFs of $|\delta\rho_r|$ are non-Gaussian and have a
longer tail than the lognormal distribution.

%fig2
\begin{figure}[htb]
\center
\includegraphics[scale=1,height=3in]{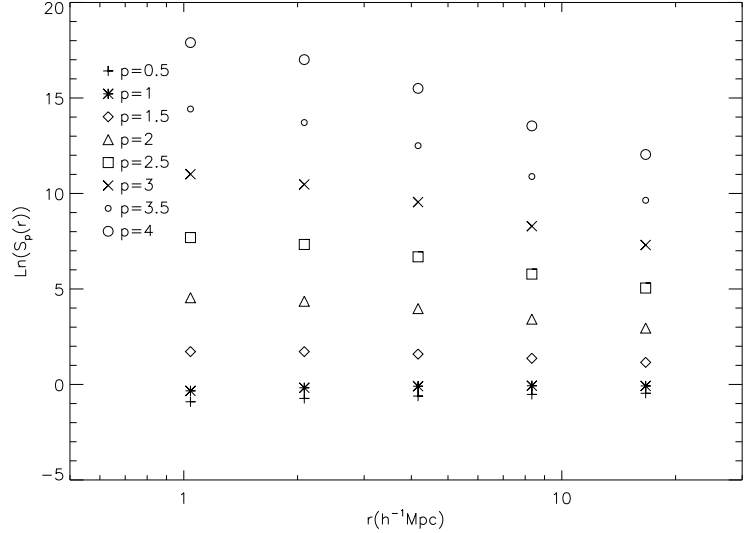}
\caption{Structure functions $S_p(r)$ vs. $r$ in the range $1 < r<
16$ h$^{-1}$ Mpc. $p$ is equal to $0.5\times n$ and $n=1,2...8$ from
bottom to top.}
\end{figure}

The long tailed events can be effectively described by the
$p$-dependence of the structure function $S_p(r)$. Figure 2 shows
$S_p(r)$ as a function of $r$ for $p=$ 0.5 to 4 for the cosmic
baryon fluid sample at $z=0$. For all $p$ the structure function
$\ln S_p(r)$ of fig. 2 can be well fitted by a straight line in the
scale range of $2\leq r \leq 16 $ h$^{-1}$ Mpc. The data points at 1
h$^{-1}$ Mpc are slightly deviating from the straight line given by
the fitting over $2\leq r \leq 16 $ h$^{-1}$ Mpc, because 1 h$^{-1}$
Mpc is already close to the Jeans length. We only focus on the range
of $2 \leq r \leq 16 $ h$^{-1}$ Mpc below.
The upper limit $16$ h$^{-1}$ Mpc actually is from the
finite size of the simulation box.

For $p>1$, the structure function decreases when the scale $r$
increases from 2 to 16 h$^{-1}$ Mpc, while for $p<1$ it increases
with the increase of scale $r$. This requires $\xi(p)<0$ for $p>1$,
and $\xi(p)>0$ for $p<1$. Therefore, the intermittent exponent can't
be fitted by $\xi(p) \propto p$. This is once again to show that the
field is highly non-Gaussian. On the other hand, equation (7) does
show that $\xi(p)$ can have different signs for $p>1$ and $p<1$.

\subsection{Hierarchical relation and parameter $\beta$}

We now test the hierarchical relation (11), which can be rewritten
as
%eq13
\begin{equation}
\ln F_{p+1}(r) =\beta \ln F_{p}(r) + A(r),
\end{equation}
where $A(r)=(1-\beta)\ln F_{\infty}(r)$ depends only on $r$.
Quantity $A(r)$ may also depend on $p$ if the proportional
coefficient of relation (9) is $p$-dependent.

Equation (13) requires that $\ln F_{p+1}(r)$ vs. $\ln F_{p}(r)$
should be a straight line for a given scale $r$, and the slope
$\beta$ should be the same for all the scales $r$. Figure 3 presents
the relations of $\ln F_{p+1}(r)$ vs. $\ln F_{p}(r)$ of the samples
at $z=0 $ for the scales $r=$ 2.1, 4.2, 8.3 and 16.7 h$^{-1}$ Mpc
and for $p=$ 1, 1.5, 2, and 2.5. It shows that all the relations of
$\ln F_{p+1}$ vs. $\ln F_p$ at different $r$ can be well fitted by
straight lines with slope $\beta=0.28\pm 0.02$.

%fig3
\begin{figure}[htb]
\center
\includegraphics[scale=1,height=3in]{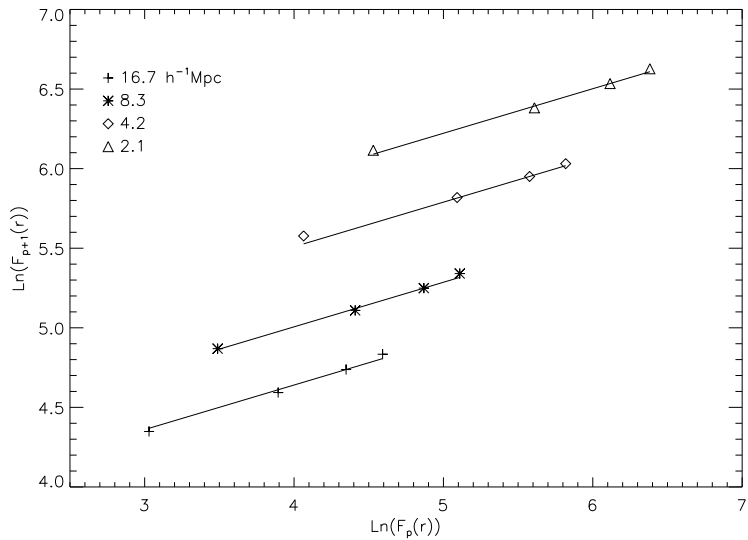}
\caption{$\ln F_{p+1}$ vs. $\ln F_p$.  $r$ is equal to
  2.1, 4.2, 8.3, and 16.7 h$^{-1}$ Mpc for lines from top to bottom.
 In each lines, the four data points correspond to $p=$ 1, 1.5, 2, and
 2.5 .}
\end{figure}

Figure 3 also shows that $A(r)$ depends only on $r$, but not on $p$.
This is consistent with the assumption of the $p$-independence of
the proportional coefficient of relation (9). If $A(r)$ is
$p$-independent, we can find the following relation from eq.(13)
%eq14
\begin{equation}
\ln [F_{p+1}(r)/F_3(r)]=\beta \ln [F_{p}(r)/F_2(r)].
\end{equation}
This is, for {\it all} $r$ and $p$, $\ln [F_{p+1}(r)/F_3(r)]$ vs.
$\ln [F_{p}(r)/F_2(r)]$ should be on a straight line. The relation
of equation (14) is tested in Figure 4. All data points can indeed
be fitted by a straight line with slope $0.28$. This is the
hierarchy of the density fluctuations between different order $p$ on
various scale $r$.

%fig4
\begin{figure}[htb]
\center
\includegraphics[scale=1,height=3in]{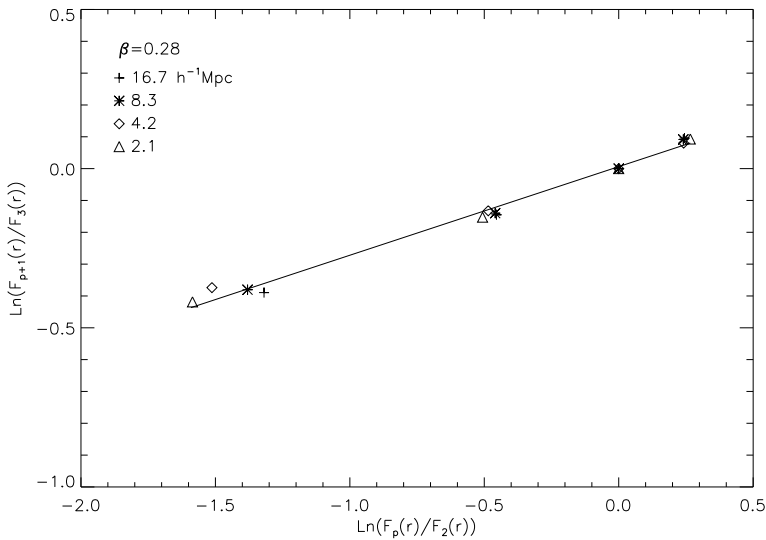}
\caption{$\ln [F_{p+1}(r)/F_3(r)]$ vs. $\ln [F_p(r)/F_2(r)]$ for
data points of $r=$2.1, 4.2, 8.3 and 16.7 h$^{-1}$ Mpc and $p=$
1, 1.5, 2, and 2.5.}
\end{figure}

\subsection{Intermittent exponent and parameter $\gamma$}

%fig5
\begin{figure}[htb]
\center
\includegraphics[scale=1,height=3in]{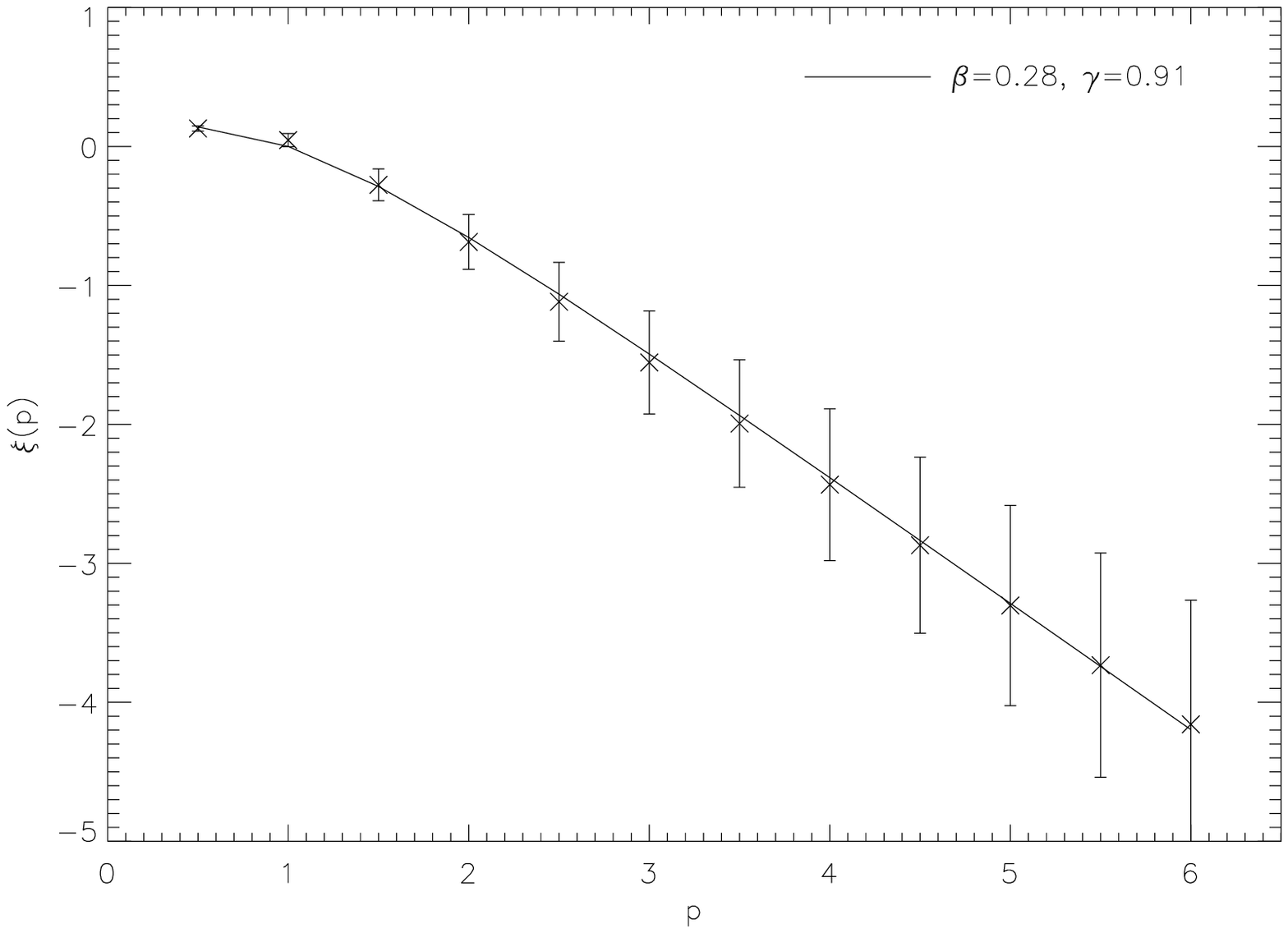}
\caption{Intermittent exponent $\xi(p)$. The solid line is given by
equation (7) with $\beta=0.28$ and $\gamma=0.91$. The data points
are from the fitting to the structure function $S_p(r)$ with
equation (7). The error bars are the variance of $\xi(p)$ over 100
samples, each of which contains of 100 one-dimensional sub-samples.
}
\end{figure}

The intermittent exponent $\xi(p)$ as a function of $p$ can be measured
by fitting $S_p(r)$ (Figure 2) with a straight line of $\ln
S_p(r)=\xi(p)\ln r +{\rm const}$ for each $p$. The measured $\xi(p)$
for the sample at $z=0$ are shown in Figure 5. The error bars are
the variance of $\xi(p)$ over 100 samples, each of which contains
100 one-dimensional sub-samples.

Equation (7) shows that the shape of $\xi(p)$ as a function of $p$
depends only on parameter $\beta$, while parameter $\gamma$ gives
the overall amplitude of the curve $\xi(p)$. Since $\beta$ is
already determined in the last section, we can determine the
parameter $\gamma$ by fitting Equation (7) to the amplitude of the
measured $\xi(p)$. The best fitting result is $\gamma=0.91$. The
fitted curve $\xi(p)$ are also shown in Figure 5. It shows that the
feature of the intermittent exponent $\xi(p)$ of the cosmic baryon
fluid at $z=0$ in the range of $0.5 \leq p \leq 6$ can be well
reproduced with the log-Poisson model with parameters $\beta=0.28$
and $\gamma=0.91$.

%fig6
\begin{figure}[htb]
\center
\includegraphics[scale=1,height=3in]{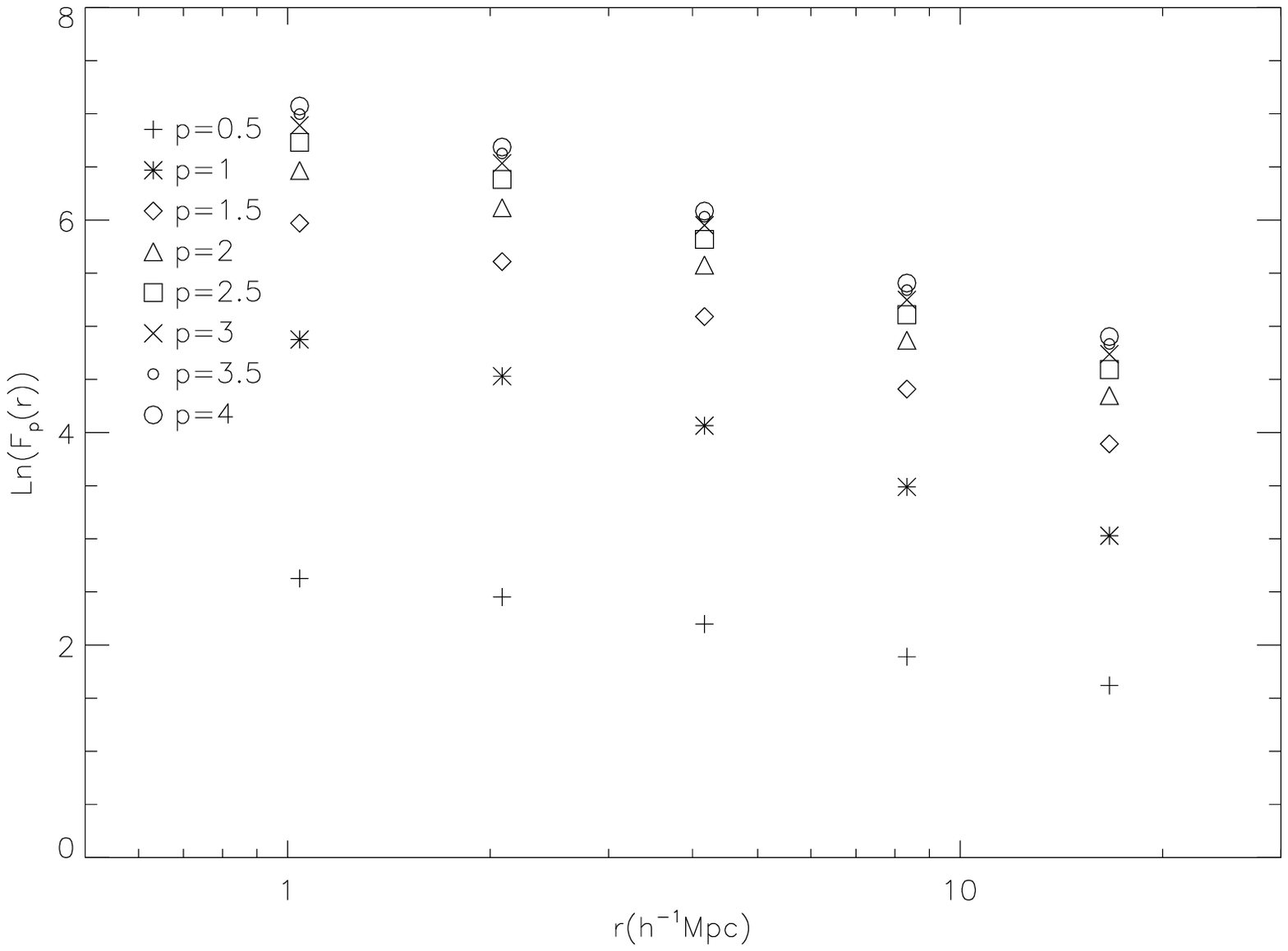}
\caption{$\ln F_p(r)$ vs. $\ln r$ for $p=0.5$ to 4 from bottom to
top.}
\end{figure}

One can make a further test of parameter $\gamma$ from equation (9).
Since $\beta=0.28$, we have $(0.28)^3 \simeq 0.02 \ll 1$, and then,
equation (9) yields
%eq15
\begin{equation}
\ln F_p (r) \simeq -\gamma \ln r + {\rm const}, \hspace{3mm} {\rm if
\ } p > 3.
\end{equation}
Equation (15) requires that the relations of $\ln F_p(r)$ vs. $\ln
r$ should be straight lines for all orders $p > 3$ with the same
slope of $\gamma$. Figure 6 presents the relation between $\ln
F_p(r)$ and $\ln r$, which can be fitted by straight lines in the
scale range of $2 \leq r\leq 16$ h$^{-1}$Mpc. The slope of the
lines with $p>3$ are $0.88\pm 0.06$, consistent with the value
of $\gamma=0.91$ determined from $\xi(p)$ (Figure 5).

\subsection{Moments}

With the determined parameters $\beta$ and $\gamma$, we can predict
statistical properties of the cosmic baryon fluid without other free
parameters. As the first one, we consider the ratio between the high
order and 2nd order moments,
$\langle\delta\rho_r^{2p}\rangle/\langle\delta \rho_r^2\rangle^p$.
When $p=2$, it is kurtosis, which is a popular tool to detect
non-Gaussianity. For a Gaussian field, the ratio should be a
constant, independent of $r$. The 2nd moment $\langle\delta
\rho_r^2\rangle$ actually is the power spectrum of the mass density
field (Fang \& Feng 2000). For the log-Poisson model we have (see
Appendix)
%eq16
\begin{equation}
\ln \frac{\langle \delta\rho_r^{2p}\rangle}{\langle
\delta\rho_r^{2}\rangle^p}= K_p\ln r + {\rm const}
\end{equation}
with
%eq17
\begin{equation}
K_p=-\gamma\frac{p(1-\beta^2)-(1-\beta^{2p})}{1-\beta}.
\end{equation}
That is, $\ln (\langle\delta\rho_r^{2p}\rangle/\langle\delta
\rho_r^2\rangle^p)$ is linearly dependent on $\ln r$ (scale free)
with the coefficient $K_p$ determined by $\beta$ and $\gamma$. As
expected, for Gaussian field $(\beta\rightarrow 1)$, $K_p=0$, i.e.,
the ratio of moments is independent on $\ln r$.

%fig7
\begin{figure}[htb]
\center
\includegraphics[scale=1,height=3in]{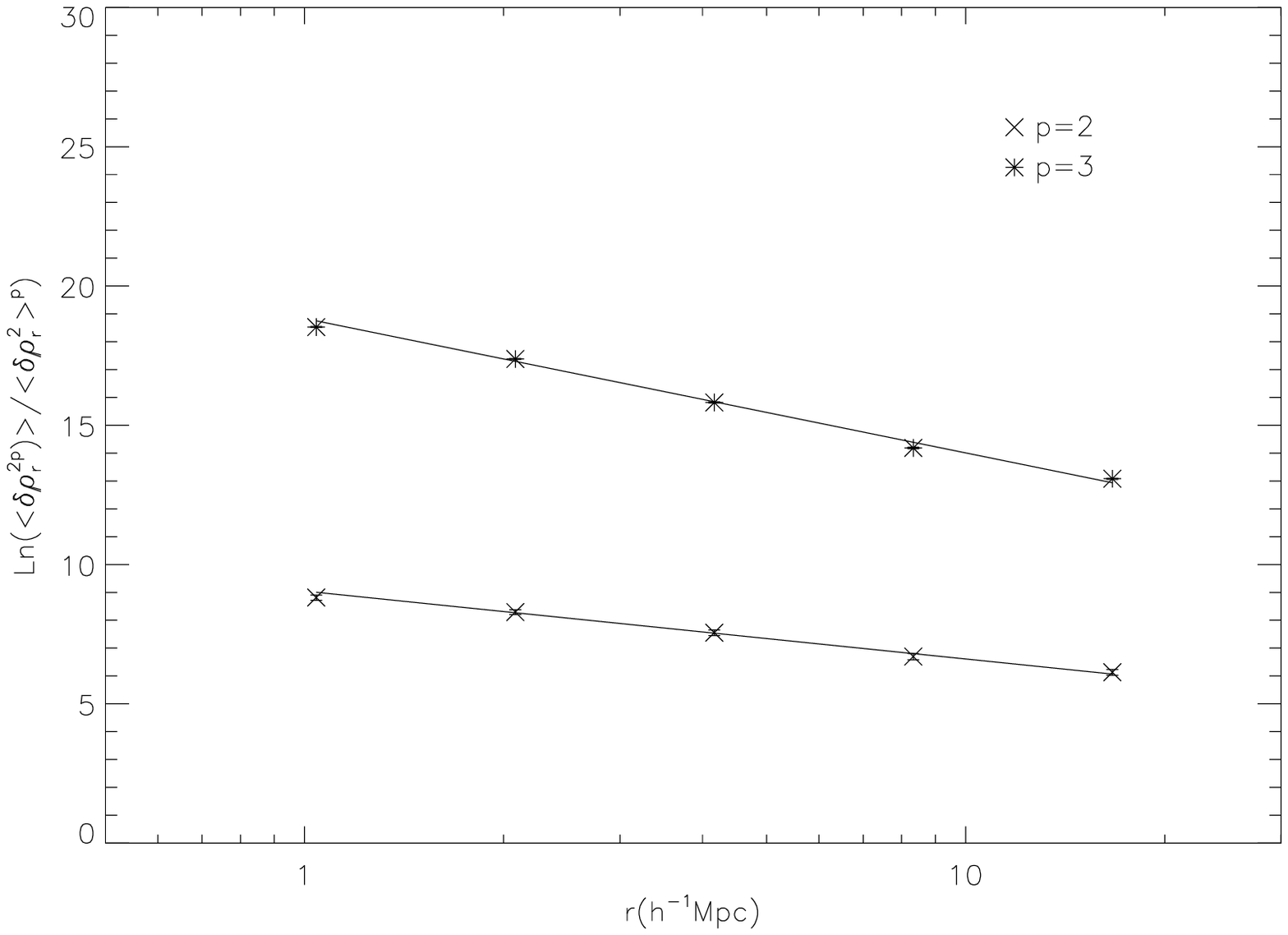}
\caption{The ratio between high order and 2nd order moments
$\langle\delta\rho_r^{2p}\rangle/\langle\delta\rho_r^{2}\rangle^p$
as a function of $r$ for $p=2$ and 3. The solid lines are given by the
least square fitting, which yields slopes consistent with the ones
calculated from equation (17) with $\beta=0.28$ and $\gamma=0.91$. }
\end{figure}

Figure 7 shows the relation of $\ln (\langle
\delta\rho_r^{2p}\rangle/\langle \delta\rho_r^{2}\rangle^p)$ vs.
$\ln r$ for the sample at redshift $z=0$. For clarity, we show only
the results of $p=$ 2 and 3, which correspond to the statistical
order 4 and 6. The errors are calculated as the variance over 100
samples, each of which contains 100 lines.
Since the error bars actually are very small in
logarithm scale, one can not show them in Figure 7. The solid lines of
Figure 7 are given by a least square fitting and have slopes $1.06\pm
0.06$ and $2.10\pm 0.12$, which are in agreement with the values
1.07 and 2.23 calculated from equation (17) with $\beta=0.28$ and
$\gamma=0.91$.

\subsection{Scale-scale correlation}

A powerful non-Gaussian detector is the so-called scale-scale
correlation, which is defined as
%eq18
\begin{equation}
C^{p,p}_{r_1,r_2}\equiv \frac{\langle \delta \rho_{r_1}^{p}\delta
\rho_{r_2}^{p}\rangle} {\langle \delta \rho_{r_1}^{p}\rangle
\langle\delta \rho_{r_2}^{p}\rangle}.
\end{equation}
Obviously, for a Gaussian field, $C^{p,p}_{r_1,r_2}=1$. It has been
shown that one can construct a non-Gaussian field, which has
identical first and second order statistics as a Gaussian field, but
 has strong scale-scale correlation (Pando et al 1998).

%fig8
\begin{figure}[htb]
\center
\includegraphics[scale=1,height=3in]{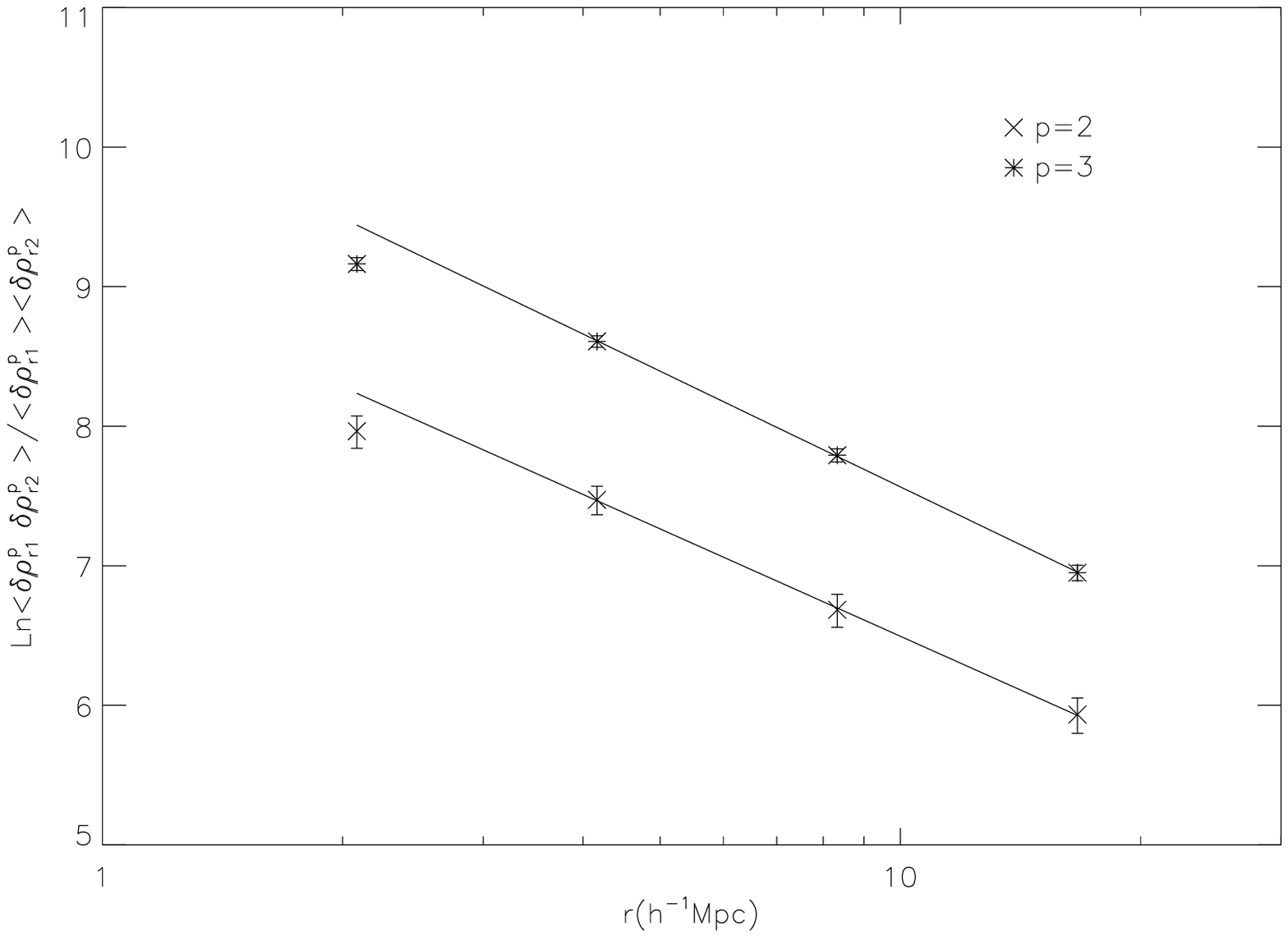}
\caption{Scale-scale correlation of the sample at $z=0$ for $p=2$
and 3 with $r_2/r_1=2$.}
The error bars are the variance over 100 samples, each of which 
contains of 100 lines.
\end{figure}

It is especially important to measure the scale-scale correlation of
cosmic baryon matter. The clustering of cosmic large scale structure
in the nonlinear regime essentially is due to the interaction
between Fourier modes on different scales (e.g., Peebles 1980).
Therefore, cosmic clustering will definitely yield the transfer of
the density perturbation powers between different scales, which
leads to scale-scale correlation. Scale-scale correlation is also
effective to distinguish various hierarchy cascade models (Pando et
al. 1998). For instance, a Gaussian hierarchy cascade, like the
model of Cole and Kaiser (1987), still yields $C^{p,p}_{r_1,r_2}=1$,
while the so-called $p$-model and $\alpha$-model yield
$C^{p,p}_{r_1,r_2}$ depending on both $p$ and $r$ (Greiner et al.
1996).

If the ratio $r_2/r_1$ is fixed, the log-Poisson model predicts the
scale-scale correlation to be (see Appendix)
%eq19
\begin{equation}
C^{p,p}_{r_1,r_2}=B(r_2/r_1)r_1^{\xi(2p)-2\xi(p)},
\end{equation}
where the coefficient $B(r_2/r_1)$ depends only on the ratio
$r_2/r_1$, as the log-Poisson model is invariant of the dilation.

Thus, assuming $r_2/r_1$ remains constant, the relationship of $\ln
C^{p,p}_{r_1,r_2}$ vs. $\ln r_1$ should be a straight line with the
slope  $\xi(2p)-2\xi(p)= -\gamma(1-\beta^p)^2/(1-\beta)$. The result
is shown in Figure 8, in which we take $r_1/r_2=2$ and $p=2$, 3.
When the number $p=3$, the statistics of equation (19) actually is
of $2p=6$ order. Figure 8 shows that for $r>$ 4$h^{-1}$Mpc, the
scale-scale correlations can be well fitted by straight lines with
slopes $1.11\pm 0.02$ and $1.20\pm 0.01$  for $p=2$ and 3, which are
consistent with the values 1.10 and 1.20 calculated from eq.(19) 
	with $\beta=0.28$
and $\gamma=0.91$. The lower limit 4 $h^{-1}$ Mpc 
for the scale-scale correlation is higher than the lower limit
2 h$^{-1}$ Mpc of the statistics in previous sections. It is because
we take $r_2/r_1=2$, and the scale-scale correlation of 4 h$^{-1}$ Mpc
actually is the correlation between modes of 4 and 2 h$^{-1}$ Mpc.

\section{Evolution of $\beta$ and $\gamma$}

%\begin{center}
\begin{table}
\begin{center}
\caption{$\beta$ and $\gamma$ at different redshift z}
\bigskip
\begin{tabular}{l|lllll}
    \tableline
    z           & 0    & 1    & 2    & 3    & 4      \\
        \tableline
    $\beta$ & 0.28 & 0.34 & 0.34 & 0.38 & 0.43    \\
    $\gamma$& 0.91 & 0.91 & 1.0 & 1.06  & 1.16   \\
    \tableline
\end{tabular}
\end{center}
\end{table}
%\end{center}

We repeat the similar analysis for samples at redshifts $z=1$, 2, 3,
and 4. The non-Gaussian features of all these samples can also be
well explained with the log-Poisson model. The parameters $\beta$
and $\gamma$ are listed in Table 1, which shows that both the
parameters $\beta$ and $\gamma$ are increasing with redshift. The
increase of $\beta$ with redshift indicates that the intermittency
is stronger at lower redshifts, and the fields at higher redshifts
contain less singular structures than that at lower redshifts.

On the other hand, the increase of $\gamma$ with redshift indicates
that the singular feature is even stronger at higher redshift. This
probably is because the baryon fluid is significantly heated by the
Burgers' shocks at lower redshift (He et al. 2004) and leads to
weaker singular structures.

\section{Discussion and conclusion}

In the nonlinear regime of cosmic clustering, the dynamical
behaviors of either dark matter or baryon fluid are complicated.
Nevertheless, it is generally believed that the evolution should be
scale-covariant in the range where the dynamical equations and
initial perturbations are scale-free. Therefore, hierarchical and
universal scaling relations have been widely used to describe the
nonlinear clustering. For instance, the hierarchical relations of
irreducible correlation functions and the universal density profile
of halos are successful in the description of the statistical
features of massive halos of dark matter.

However, it has already been recognized in the early study of cosmic
structure formation that in the nonlinear regime the dynamical behavior
of cosmic baryon doesn't always follow the collisionless
dark matter. The non-Gaussianity of the mass and velocity fields of
baryon fluid cannot be given by a similar mapping of the mass and
velocity fields of dark matter. For instance, the halo model assumes that
all mass fields are given by a superposition of the halos on
various scales, and all non-Gaussian behaviors of the density field
are described by the universal density profile (e.g., Cooray \& Sheth 2002),
this makes it difficult to explain the intermittency and the scale-scale
correlation of the transmitted flux in the absorption spectra of QSOs.

We show that the evolution of the cosmic baryon fluid, governed by the
Navier-Stokes equation in an expanding universe, is also hierarchical
in the scale range in which the dynamical equations and initial
perturbations are scale-free. The non-Gaussian behavior of the
mass density field of baryon fluid can be well explained by
the log-Poisson hierarchical cascade model. The SL formula and/or
log-Poisson model are universal for the fully developed turbulence
of Navier-Stokes fluid in the scale-free range. Therefore, the result
of this paper implies that, in the scale-free range, the cosmic baryon
fluid reaches a statistically quasi-steady state. For a fully developed
turbulence, energy passes from large to the smallest eddies and
finally dissipates into thermal motion, while the cosmic baryon fluid
undergoes the evolution of clustering and finally falls into
massive halos of dark matter to form structures, including light-emitting
objects. Therefore, the log-Poisson model works on the scale range
from the onset scale of the nonlinear evolution (a few tens of h$^{-1}$
Mpc) to the dissipation scale, i.e., the Jeans length.

In view of this picture, one can say that in the nonlinear regime,
the statistical properties of the cosmic baryon fluid are actually
less dependent on the details of the dissipative processes. This
property has already been noted in describing baryon matter by the
Burgers' equation. Although the Burgers' equation contains a
dissipative term, which leads to the formation of shocks and
condense into luminous objects (Jones 1999), the self-similar
properties of Burgers' turbulence actually depend very weakly
on the dissipative term.

We now address the possible applications of the log-Poisson model.
First, it is interesting to compare the log-Poisson model with the
lognormal model, which assumes that the PDFs of the cosmic baryon
matter is log-normal and no details of dissipative processes are
needed (Bi \& Davidsen 1997). The lognormal model is successful to
explain some statistical features of Ly$\alpha$ forests and also
predicts that the transmitted flux in the spectra of QSOs is
non-Gaussian and intermittent. This result is qualitatively
consistent with the observed data; however, non-Gaussian features
given by the lognormal model do not quantitatively fit the observed
data. For instance, the high order moment (\S 3.4) given by the
lognormal model has a $K_p \propto p(p-1)$, while the data show
$K_p\propto -p^{0.1}(p-1)$ (Pando et al. 2002). The later is
actually close to the log-Poisson model. Therefore, the higher order
statistics of the Ly$\alpha$ transmitted flux would be able to
discriminate between the log-Poisson and the lognormal model.

Second, recent studies have shown that the turbulence behavior of
baryon gas can be detected by the Doppler-broadened spectral lines
(Sunyaev et al. 2003; Lazarian \& Pogosyan 2006). Although these
works focus on the turbulence of baryon gas in clusters, the result
is still applicable, at least, for the warm-hot intergalactic medium
(WHIM), which is shown to follow the  evolution of Burgers' fluid on
large scales (He et al. 2004, 2005). The last but not least, the
polarization of CMB is dependent on the density of electrons, and
therefore, the map of CMB polarization would provide a direct test
on the non-Gaussian features of ionized gas when the data on small
scales becomes available.

\acknowledgments J.L acknowledges the financial support of the
International Center for Relativistic Astrophysics. This work is
supported in part by the US NSF under the grant AST-0507340. We also
thank Mr. Ding Ma for his contributions in the early stage of this
project.

\appendix

\section{Log-Poisson model and intermittancy exponent}

In this appendix we give the details of deriving the statistical
properties of the log-Poisson cascade model. Let us
consider the log-Poisson model
%eqA1
\begin{equation}
|\delta\rho_{r_1}| = W_{r_0r_1}|\delta\rho_{r_0}|,
\end{equation}
where
%eqA2
\begin{equation}
W_{r_0r_1}=\beta^m (r_0/r_1)^{\gamma},
\end{equation}
and $m$ is a Poisson variables with probablity distribution function
%eqA3
\begin{equation}
P(m)=\exp(-\lambda_{r_0r_1})\lambda_{r_0r_1}^m/m!.
\end{equation}
and
%eqA4
\begin{equation}
\lambda_{r_0r_1}= \gamma[\ln(r_0/r_1)]/(1-\beta).
\end{equation}
Thus
%eqA5
\begin{eqnarray}
\langle W_{r_0r_1}^p \rangle & = & \sum_m (\beta^m (r_0/r_1)^{\gamma})^p
\exp(-\lambda_{r_0r_1})\lambda_{r_0r_1}^m/m! \\
  &=& \exp(-\lambda_{r_0r_1})(r_0/r_1)^{{\gamma}p}
    \sum_m \beta^{mp}\lambda_{r_0r_1}^m/m! \\
  &=& e^{-\lambda_{r_0r_1}}e^{{\gamma}p\ln(r_0/r_1)}
    \sum_m (\beta^p\lambda_{r_1r_2})^m/m! \\
  &=& e^{-\lambda_{r_0r_1}}e^{{\gamma}p\ln(r_0/r_1)}
      e^{\beta^p\lambda_{r_0r_1}} \\
\end{eqnarray}

Using equation (A4), we have
%eqA6
\begin{equation}
\langle W_{r_0r_1}^p \rangle= (r_0/r_1)^{-\xi(p)},
\end{equation}
with
%eqA7
\begin{equation}
\xi(p)=- \gamma[p - (1-\beta^p)/(1-\beta)].
\end{equation}
Therefore
%eqA8
\begin{equation}
\frac{S_p(r_1)}{S_p(r_2)}=
\frac{\langle W_{r_0r_1}^p \rangle}
{\langle W_{r_0r_2}^p \rangle}= \left(\frac{r_1}{r_2}\right )^{\xi(p)}.
\end{equation}

For moments equation, we have
%eqA9
\begin{equation}
\frac{\langle \delta\rho_r^{2p}\rangle}{\langle
\delta\rho_r^{2}\rangle^p}=
\frac{\langle W_{r_0r}^{2p} \rangle}{(\langle W_{r_0r}^2 \rangle)^p}
=(r/r_0)^{\xi(2p)-p\xi(2)}.
\end{equation}
Therefore,
%eqA10
\begin{equation}
\ln \frac{\langle \delta\rho_r^{2p}\rangle}{\langle
\delta\rho_r^{2}\rangle^p}= K_p \ln r + {\rm const},
\end{equation}
and
%eqA11
\begin{equation}
K_p=\xi(2p)- p\xi(2)=-\gamma\frac{p(1-\beta^2)-(1-\beta^{2p})}{1-\beta}.
\end{equation}

For scale-scale correlation, we have
%eqA12
\begin{equation}
C^{p,p}_{r_1,r_2}=\frac{\langle \delta \rho_{r_1}^{p}\delta
\rho_{r_2}^{p}\rangle} {\langle \delta \rho_{r_1}^{p}\rangle
\langle\delta \rho_{r_2}^{p}\rangle}=
\frac{\langle W_{r_0r_1}^{p}W_{r_0r_2}^{p}\rangle}
 {\langle W_{r_0r_1}^{p}\rangle \langle W_{r_0r_2}^{p}\rangle}.
\end{equation}
Using equation (A10), if keeping $r_1/r_2$ to be
constant, the $r_1$-dependence of $C^{p,p}_{r_1,r_2}$ is given by
%eqA13
\begin{equation}
C^{p,p}_{r_1,r_2}=A(r_2/r_1)r_1^{\xi(2p)-2\xi(p)}.
\end{equation}

\end{document}